\pdfoutput=1
\documentclass[aps,prl,reprint,amsmath,amssymb,superscriptaddress]{revtex4-1}
\usepackage{bm}
\usepackage{graphicx}
\usepackage[colorlinks]{hyperref}
\usepackage{mathrsfs}
\usepackage{times}
\usepackage[dvipsnames]{xcolor}
\usepackage{float}
\usepackage{braket}

\newcommand{\bit}{\begin{enumerate}}
	\newcommand{\eit}{\end{enumerate}}

\definecolor{bananayellow}{rgb}{1.0, 0.88, 0.21}
\definecolor{straw}{rgb}{0.32, 0.28, 0.1}

\begin{document}
	\title{Localization persisting under {aperiodic} driving}
		\author{Hongzheng Zhao}
\affiliation{\small Max-Planck-Institut f{\"u}r Physik komplexer Systeme, N{\"o}thnitzer Stra{\ss}e 38, 01187 Dresden, Germany}
	\affiliation{\small Blackett Laboratory, Imperial College London, London SW7 2AZ, United Kingdom}
	
	\author{Florian Mintert}
	\affiliation{\small Blackett Laboratory, Imperial College London, London SW7 2AZ, United Kingdom}
		\author{Johannes Knolle  }
	\affiliation{Department of Physics TQM, Technische Universit{\"a}t M{\"u}nchen, James-Franck-Stra{\ss}e 1, D-85748 Garching, Germany}
	\affiliation{Munich Center for Quantum Science and Technology (MCQST), 80799 Munich, Germany}
	\affiliation{\small Blackett Laboratory, Imperial College London, London SW7 2AZ, United Kingdom}
	
	\author{Roderich Moessner}
		\affiliation{\small Max-Planck-Institut f{\"u}r Physik komplexer Systeme, N{\"o}thnitzer Stra{\ss}e 38, 01187 Dresden, Germany}
	
	\begin{abstract}
\end{abstract}

\begin{abstract}
Localization may survive in periodically driven (Floquet) quantum systems, but is generally unstable for aperiodic drives. In this work, we identify a hidden conservation law originating from a chiral symmetry in a disordered spin-$\frac{1}{2}$ XX chain. This protects indefinitely long-lived localization for general--even aperiodic--drives.
Therefore, rather  counter-intuitively, adding further potential disorder which spoils the conservation law {\it delocalizes} the system, via a controllable parametrically long-lived prethermal regime. This provides a first example of persistent single-particle `localization without eigenstates'.
\end{abstract}
	\maketitle

\textit{Introduction.--} 
Closed quantum many-body systems with time-dependent (`driven') Hamiltonians tend to absorb energy until they reach  a featureless `infinite temperature' state~\cite{2014PhRvE..90a2110L,abanin2015exponentially,kuwahara2016}. This heat death can be prevented if a sufficient number of integrals of motion is present. This may occur in integrable systems~\cite{PGE}, or via the emergent structure underpinning many-body localization~\cite{abanin2017recent}, which is stable against periodic driving for sufficiently strong disorder~\cite{lazarides2015fate,Ponte2015}. This even allows for novel non-equilibrium  phenomena such as discrete time crystalline order~\cite{Khemani_TC,FTC} and anomalous Floquet-Anderson insulators~\cite{titum2016anomalous}.

In recent years,
the focus has shifted to aperiodic drives to control quantum systems~\cite{Verdeny2016,dumitrescu2018logarithmically,else2020,long2021many,martin2017topological,crowley2019topological,korber2020interacting,boyers2020exploring,dumitrescu2018logarithmically,Zhao2019,else2020,nandy2017aperiodically,lapierre2020fine,wen2021periodically}. This leads to new types of dynamical phenomena, e.g. non-adiabatic topological energy pumps~\cite{long2021nonadiabatic,nathan2021quasiperiodic,qi2021universal} and prethermalization with algebraically scaling lifetimes~\cite{zhao2021random,mori2021rigorous}. 
 However, the absence of time translation symmetry (TTS) is generally expected to generate stronger heating effects, in particular destabilizing localization~\cite{dumitrescu2018logarithmically,else2020}. Nonetheless, localization has been shown to remain robust for two-tone continuous drives~\cite{long2021many}, paving the way to stabilize non-equilibrium phases unobtainable with periodic drives, {\it e.g.}, the quantum Hall effect in synthetic dimension~\cite{martin2017topological,crowley2019topological,korber2020interacting,boyers2020exploring} and discrete-time quasicrystals~\cite{dumitrescu2018logarithmically,Zhao2019,else2020}. 
 
 One natural question, then, is under what conditions localization can indefinitely survive
 generic aperiodic and even random driving profiles? Unlike multi-tone drives with few frequency components~\cite{long2021many,dumitrescu2018logarithmically}, generic drives contain low-frequency components which normally induce delocalization.
 Technically (and conceptually), the analysis of generic aperiodic drives is complicated as no way is known of analysing the quantum dynamics in terms of a fixed set of eigenstates, $\{\psi_i\}$, whose time evolution is described by a simple phase accumulation rate given by their respective (quasi-)energies $\epsilon_i$. 
Here, we show that a localized steady state can emerge in a non-interacting multi-particle system
subjected to a drive with a well-defined coupling operator but general time-dependence
and in this sense represents single-particle localization without eigenstates.

Our model hosts disorder not as an on-site potential but instead via randomness in the exchange interactions (or hopping amplitudes for fermions)~\cite{PHSMBL,de2020anomalous,fisher1994random,de2020anomalous}. The disordered exchange interaction has a single-particle chiral symmetry resulting in a spectrum symmetric around zero. Unlike the single-particle localization induced by potential disorder, where all eigenstates are localized,
the chiral symmetric randomness does not fully localize all eigenstates, but instead yields a diverging localization length at zero energy~\cite{Dyson,Soukoulis1981,fleishman1977fluctuations,de2016generalized}. Crucially, this chiral symmetry enables us to construct drives  with a hidden conservation law, and we demonstrate that the localized steady state persists indefinitely for arbitrary aperiodic drives as long as the conservation law is preserved.

A static  disorder-free exchange without the chiral symmetry, or potential disorder, destroys this feature. They do so via different routes, as the latter counterintuitively combines the {\it addition} of disorder with {\it de}localization as the symmetry is broken. This `competition' gives rise, as a function of disorder strength, to a non-monotonic heating rate. The resultant `prethermal' localized state can also be exceptionally  long-lived upon increasing the driving frequency.

In the following, we first introduce the model with the single-particle chiral symmetry, and show that the system reduces to a collection of two-level systems. Each pair corresponds to the localized chiral symmetric excitations at opposite energies $\pm \epsilon_i$. We then construct a drive which only introduces dynamics within each two-level system. Therefore, single-particle localization always persists despite the absence of well-defined eigenstates for generic aperiodic drives.
For a multi-particle state, however, the single-particle chiral symmetry does not straightforwardly generalize. Instead, we show that it leads to an extensive number of local conserved quantities which can protect the multi-particle steady state from delocalization. 

We then focus on the aperiodic Thue-Morse (TM) drive to numerically verify the existence of persistent localization via analyzing both the single-particle time evolution operator and the entanglement generation for multi-particle states. We finally show how different types of chiral symmetry-breaking perturbations lead to a long-lived prethermal regime with {\it non-monotonic} scalings of the prethermal lifetime.

\textit{The model and hidden conservation law.--} We focus on the spin-$1/2$ model described by the Hamiltonian 
\begin{eqnarray}
H(t) =H_0+{\delta h}f(t)\hat{P}, 
\label{eq.Hamiltonian}
\end{eqnarray}
where the nearest-neighbor exchange interaction reads $H_0 = \frac{1}{4}\sum_{i}J_i\big(\sigma_i^x\sigma_{i+1}^x+\sigma_i^{y}\sigma_{i+1}^y\big)$ with disordered coupling strength $ J_i\in[J_0,J_0+{J_{max}}]$. The staggered potential $\hat{P} = \sum_i(-1)^i\sigma_i^z/2$ is modulated with the amplitude $\delta h$ according to a time-dependent function $f(t)$ which will be specified later.
Since $[H_0,\hat{P}]\neq0$, the drive is not trivially equivalent to a quench. This model can be transformed into a non-interacting fermionic form (Eq.~\ref{eq.fermi}) via the Jordan-Wigner transformation~\cite{nielsen2005fermionic}
\begin{eqnarray}
H(t)= \sum_{i} \frac{J_i}{2}(c_i^{\dagger}c_{i+1}+h.c.)+ \delta hf(t)\sum_{i} (-1)^in_i,
\label{eq.fermi}
\end{eqnarray}
with the standard fermonic operators $c_i^{(\dagger)}$ and the number operator $n_i$.
The model has a $U(1)$ symmetry, hence,
the total particle number, or the total magnetization in  spin language, is conserved. As the nearest-neighbor hopping is disordered, eigenstates of $H_0$ are localized (or quasi-localized if the corresponding eigenenergies are close to zero~\cite{asboth2016schrieffer,ryu2010topological}). 
As shown in the following, the Hamiltonian $H_0$ has a hidden conservation law even in the presence of $\hat P$ which can  protect the system from delocalizing regardless of the driving profile $f(t)$.

This conservation law originates from a single particle chiral symmetry of the Hamiltonian $H_0$.
Consider a general real quadratic Hamiltonian $H_{\mathrm{C}}=\sum_{i j} c_{i}^{\dagger} A_{i j} c_{j}$, its anti-commutator with the staggered potential $\hat{P} = \sum_i(-1)^in_i$ reads
\begin{eqnarray}
\label{eq.anticommutator}
\{H_{\mathrm{C}}, \hat{P}\}&=\sum_{i j}\left[(-1)^{i}+(-1)^{j}\right] A_{i j} c_{i}^{\dagger} c_{j}+\hat{D}.
\end{eqnarray}
with $\hat{D}=2 \sum_{i jk} (-1)^{k}A_{i j} c_{i}^{\dagger}n_{k} c_{j}$.
Crucially, the matrix elements of $\hat{D}$ can be non-zero only for multi-particle states, but they vanish in the single-particle subspace.
In addition, for the Hamiltonian $H_0$ considered in Eq.~\ref{eq.Hamiltonian}, the first contribution on the right hand side of Eq.~\ref{eq.anticommutator} vanishes because $A_{ij}$ is nonzero only for $j=i+1$; in this case, however, the corresponding prefactor $(-1)^{i}+(-1)^{j}$ vanishes.
The anti-commutator $\{H_{\mathrm{0}}, \hat{P}\}$ thus vanishes in the single-particle subspace, which implies that
the Hamiltonian $H_0$ is single-particle chiral symmetric: for a given single-particle eigenstate $\ket{\epsilon_k}$ of eigenenergy $\epsilon_k$, $\hat{P}\ket{\epsilon_k}$ is also an eigenstate of opposite energy $-\epsilon_k$. 
 
The Hamiltonian $H_0$ can thus be formally diagonalized as $H_0=\sum_{0<k\leq L/2} \epsilon_{k} \gamma_{k}^{\dagger} \gamma_{k}-\epsilon_{k} \tilde{\gamma}_{k}^{\dagger} \tilde{\gamma}_{k}$ with a new set of fermionic operators 
$\gamma_k,\tilde{\gamma}_k$ defined as
\begin{eqnarray}
\label{eq.PHSstate}
\gamma_{k}^{\dagger}=\sum_{i} V_{k i} c_{i}^{\dagger},\ \  \tilde{\gamma}_{k}^{\dagger}=\sum_{i}(-1)^{i} V_{k i} c_{i}^{\dagger},
\end{eqnarray} where the matrix $\mathbf{V}$ diagonalizes the coupling matrix $\mathbf{A}$. 
This enables the expansion of the operator $\hat{P}$ in the eigenbasis as $\hat{P}=\sum_{0<k\leq L/2} \hat{P}_k$ where $\hat{P}_k=\gamma_{k}^{\dagger} \tilde{\gamma}_{k}+\tilde{\gamma}_{k}^{\dagger} \gamma_{k}$, which exchanges the excitation in each $k-$subspace as $\hat{P}_k \gamma_{k}^{\dagger}|0\rangle=\tilde{\gamma}_{k}^{\dagger}|0\rangle $. Properties of the operator $\hat{P}$ will be further discussed in the Supplementary Material (SM)~\cite{SM}. 
 
The key point is that for our choice of $\hat{P}$ as the driving term, the chiral disordered model reduces to a collection of two level systems, {\it e.g.}, by introducing the `pseudospin' representation $\ket{\uparrow}_k = \gamma_{k}^{\dagger}\ket{0},\ket{\downarrow}_k = \tilde{\gamma}_{k}^{\dagger}\ket{0}$, the drive operator $\hat{P}_k$ acts as the $\sigma^k_x$ operator in each $k-$subspace. According to Eq.~\ref{eq.PHSstate}, the paired excitations have the same spatial wavefunction up to the $(-1)^i$ phase between neighboring sites. 
Consequently, the coupling between the paired states via $\hat{P}_k$ will not significantly change the localization properties of the time evolution operator $U(t)=\mathcal{T}\left\{\exp \left(-i \int_{0}^{t} d t^{\prime} H(t^{\prime})\right)\right\}$ in the single-particle subspace at any time $t$ for arbitrary forms of the driving profile $f(t)$. 

Crucially, this symmetry argument applies only to the single-particle subspace, but not to the multi-particle subspace,
in which Eq.~\ref{eq.anticommutator} generally does not vanish due to the presence of the operator $\hat{D}$. Although the spectrum for multiple particles remains symmetric, the operator $\hat{P}$ does not simply map between multi-particle paired eigenstates of $H_0$, but instead couples many (see SM). Nevertheless, the coupling is highly constrained as the total number of excitations in each $k-$subspace, $N_k=\gamma_{k}^{\dagger} \gamma_{k}+\tilde{\gamma}_{k}^{\dagger}\tilde{\gamma}_{k}$, is a conserved quantity! It trivially commutes with the Hamiltonian $H_0$, but it crucially also commutes with $\hat{P}$ which can only convert the excitation at a fixed $k$ but cannot mix different $k$ subspaces~\cite{SM}. 
The extensive number of conserved quantities $N_k$, which are local because of the localized single-particle spectrum of $H_0$, can protect many-particle states from delocalization. 

This conservation law can be broken by single-particle chiral symmetry-breaking perturbations, {\it e.g.}, static on-site disorder $\sum_ih_in_i$ with $h_i$ randomly chosen from $[0,h_{max}]$. It mixes different $k-$subspaces, hence, eigenstates of the time evolution operator $U(t)$ may become delocalized.
However, as we will illustrate in the following, if $f(t)$ is periodic or constant, additional potential disorder will not delocalize the system but rather enhance the localization. Delocalization only happens when $f(t)$ is aperiodic, hence, highlighting the great importance of the conservation law in protecting the localization especially when TTS is absent.

\textit{Numerical results.--}
Although when $N_k$ is conserved, the concrete form of the driving profile $f(t)$ is irrelevant for the persistent localization, here we choose for concreteness the discrete quasi-periodic Thue-Morse (TM) sequence which enables us to probe exponentialy long times~\cite{nandy2017aperiodically,zhao2021random,mori2021rigorous}. By defining two Hamiltonians 
\begin{eqnarray}
H_{\pm} =H_0+\sum_ih_in_i\pm{\delta h}\hat{P},
\end{eqnarray}
 one can construct two unitary evolution operators
$U_{\pm}=\exp \left(-i T H_{\pm}\right),$ with the characteristic time scale $T$.
One can recursively define the following unitary operators
\begin{equation}
\label{eq.recursive}
U_{n+1}=\tilde{U}_{n} U_{n}, \quad \tilde{U}_{n+1}=U_{n} \tilde{U}_{n},
\end{equation}
with $U_{1}=U_{-} U_{+}$ and $\tilde{U}_{1}=U_{+} U_{-}$. The unitary time evolution for TM drive at stroboscopic times $t_m=2^mT$ is obtained as   $|{\psi(t_m)}\rangle = U_m|{\psi(0)}\rangle$~\cite{mori2021rigorous}. 
\begin{figure}
	\centering
	\includegraphics[width=0.99\linewidth]{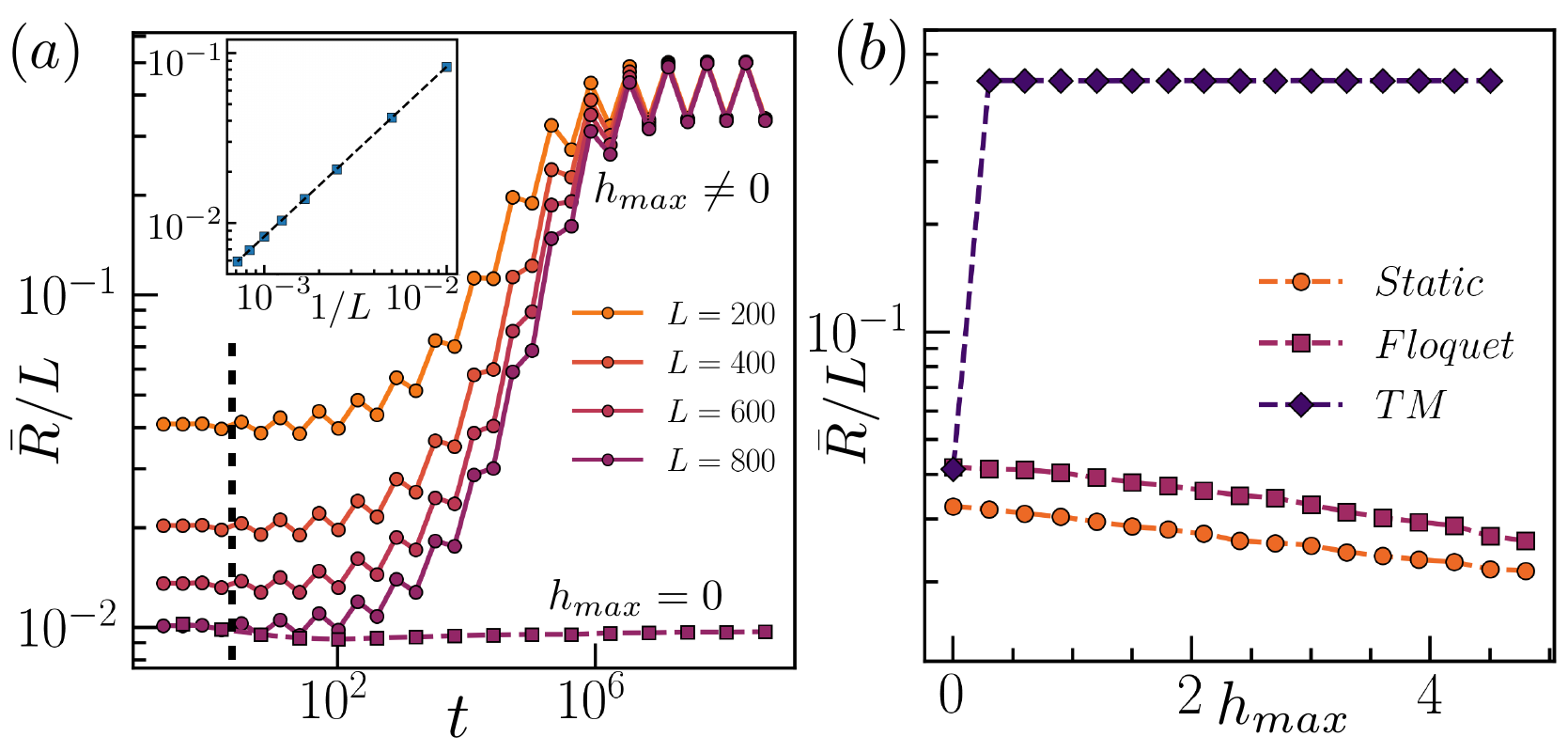}
	\caption{(a) Average PR for eigenstates of the single-particle time evolution operator $U_m$ at time $t=2^mT$. 
	Results are calculated for different system sizes and averaged over 100 disorder realization. The PR remains low in the absence of potential disorder, whereas a finite value $h_{max}=0.51$ results in delocalization. The inset shows the size dependence of $\bar{R}/L$ at $t\approx 10$ and the scaling $1/L$ is obtained to confirm the localization for short time.  (b) PR for the static Hamiltonian, Floquet operator and TM driving after sufficiently long time $t\approx 10^{12}$ computed for $L=200$. Potential disorder leads to stronger localization in static and Floquet systems, while any non-vanishing $h_{max}$ delocalizes the system driven quasi-periodically. We use $J_0=1,J_{max}=1.6\pi,\delta h=1,h_{max}=0.51,1/T=10$ for the simulation.}
	\label{fig:IPR}
	
\end{figure}
For Floquet systems, to diagnose the localization properties of single-particle eigenstates,  one can use the participation ratio (PR) defined as
$
R_k = 1/\sum_i |\psi_i^k|^4,
$
where $\psi_i^k$ denotes the $k$-th eigenfunction at site $i$ of the Floquet operator~\cite{roy2018multifractality,kramer1993localization}. To quantify the average spreading of the eigenvectors in real space, the average PR is defined over all eigenstates and different disorder realizations $\bar{R} = \langle R_k\rangle_{k,r}$. For general delocalized states, one obtains $\bar{R}/L \sim O(1)$, whereas for perfectly localized ones $\bar{R}/L \sim 1/L$~\cite{kramer1993localization}. 
However, in contrast to Floquet systems, a time-independent Floquet time evolution operator and its eigenstates are not available for our aperiodic system. Alternatively, $\bar{R}(t)/L$ is computed via the instantaneous eigenstates of the time evolution operator $U_m$ at each stroboscopic time $t=2^mT$.

As shown in Fig.~\ref{fig:IPR} (a), for the case without potential disorder $h_{max}=0$, the average PR $\bar{R}/L$ remains at a constant small value indicating that single-particle eigenstates of the evolution operator remain localized. Indeed, for each $k-$subspace, in the long-time limit, the time evolution operator recursively determined by Eq.~\ref{eq.recursive} can be directly diagonalized by $\ket{\uparrow}_k$ and $\ket{\downarrow}_k$, the same eigenstates of the static Hamiltonian $H_{\mathrm{0}}$~\cite{nandy2017aperiodically,SM}.
In contrast, the inclusion of potential disorder drastically changes the behavior as different $k-$subspaces start to mix: following a plateau at low values for finite time $t\sim 10^2$, $\bar{R}/L$ rapidly grows to system size independent values ($O(1)$) indicating delocalization. We verified that delocalization occurs for any nonzero value $h_{max}$ as shown in Fig.~\ref{fig:IPR} (b) where the average PR at a long time $t\approx 10^{12}$ is plotted. As a comparison,  for the Floquet operator $U_-U_+$ and the static Hamiltonian $H_+$, the average $\bar{R}/L$ always reduces for larger $h_{max}$ suggesting that eigenstates are more localized. Such a comparison suggests that the single-particle chiral symmetry protection becomes particularly crucial for aperiodic drives.

To verify the existence of persistent localization in multi-particle systems and to quantify the delocalization crossover at late times in the presence of a potential disorder, we next study the entanglement entropy
$
S_{L}(t) =-\operatorname{Tr}\left[\rho_{L/2} \log_2 \rho_{L/2}\right]
$
with the reduced density matrix $\rho_{L/2}$ of the half chain . 

\begin{figure}
	\centering
	\includegraphics[width=0.99\linewidth]{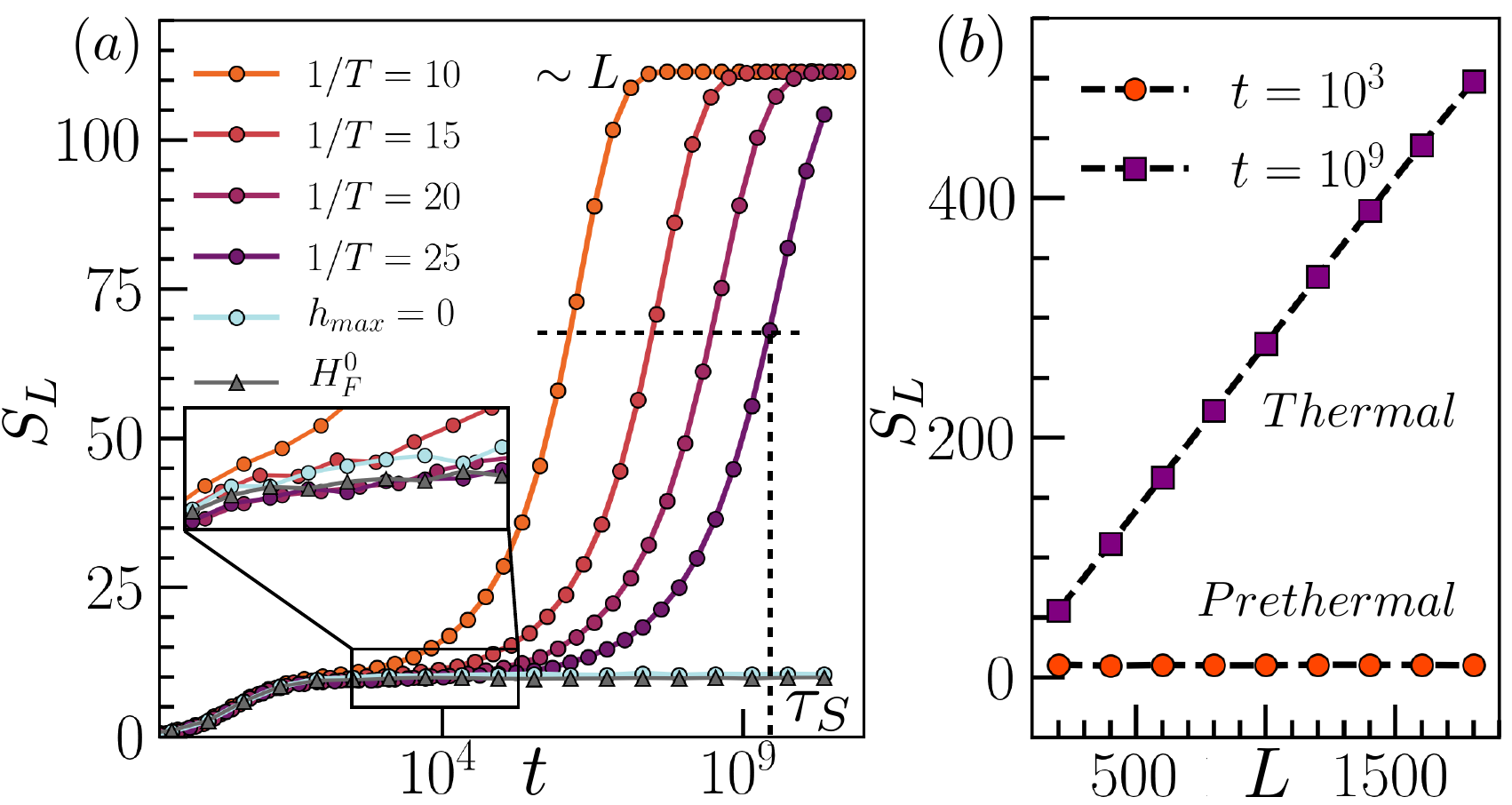}
	\caption{(a) Entanglement dynamics for different driving rates, with or without potential disorder, averaged over 100 disorder realization and different initial states. Steady state can be delocalized by a non-vanishing potential disorder and thermalize. (b) System size dependence of the entanglement entropy. Prethermal (red dots) and the final plateau (purple squares) exhibit an area- and volume-law scaling respectively. We use parameters $J_0=1,J_{max}=1.6\pi,\delta h=1,h_{max}=0.51, L=400$ for the simulation. }
	\label{fig:entanglement}
\end{figure}

The initial state is chosen as a random product state with zero entanglement in the total magnetization $S_z=0$ sector (or half-filling in the  fermionic representation). 
After switching on the drive, the system starts generating entanglement and its dynamics for varying driving rates $1/T$ is plotted in Fig.~\ref{fig:entanglement} (a).  In the absence of the potential disorder (blue), as eigenstates of $U_m$ always remain localized, entanglement can only be generated within limited regions. Therefore, it saturates around $t\approx10^2$ at a low value after a transient process, suggesting the appearance of a persistent localized steady state protected by the conservation law of $N_k$. 
Once we turn on the potential disorder, localization becomes prethermal with a finite but long lifetime $\tau_S$ for large driving rates, e.g. $1/T\geq 15$.  This prethermal localization can be well captured by an effective Hamiltonian (gray triangles), $H_{\mathrm{eff}}^0=(H_++H_-)/2$, obtained from the average of the two driving Hamiltonians~\cite{zhao2021random}.

As illustrated in the inset, the prethermal plateau (gray triangles) is slightly below the result for the steady state (blue), implying that the additional potential disorder leads to enhanced prethermal localization despite the fact that it eventually delocalizes the system at late times.

The entanglement entropy exhibits a well-pronounced increase after the time scale $\tau_S$ towards the final plateau, suggesting the eventual delocalization. As illustrated in Fig.~\ref{fig:entanglement} (b), in contrast to the prethermal entanglement which is system size independent (orange dots), the final plateau exhibits  volume-law scaling (purple squares) demonstrating that entanglement has been established extensively over the whole system. However, as the system is non-interacting, it does not heat up to infinite temperature. The final entanglement entropy is smaller than the average entropy for a random state, {\it i.e.}, the Page value $S_{\infty}=(L\log 2-1)/2$~\cite{page1993average}.

\textit{Prethermal lifetime.--}
\begin{figure}
	\centering
		\includegraphics[width=\linewidth]{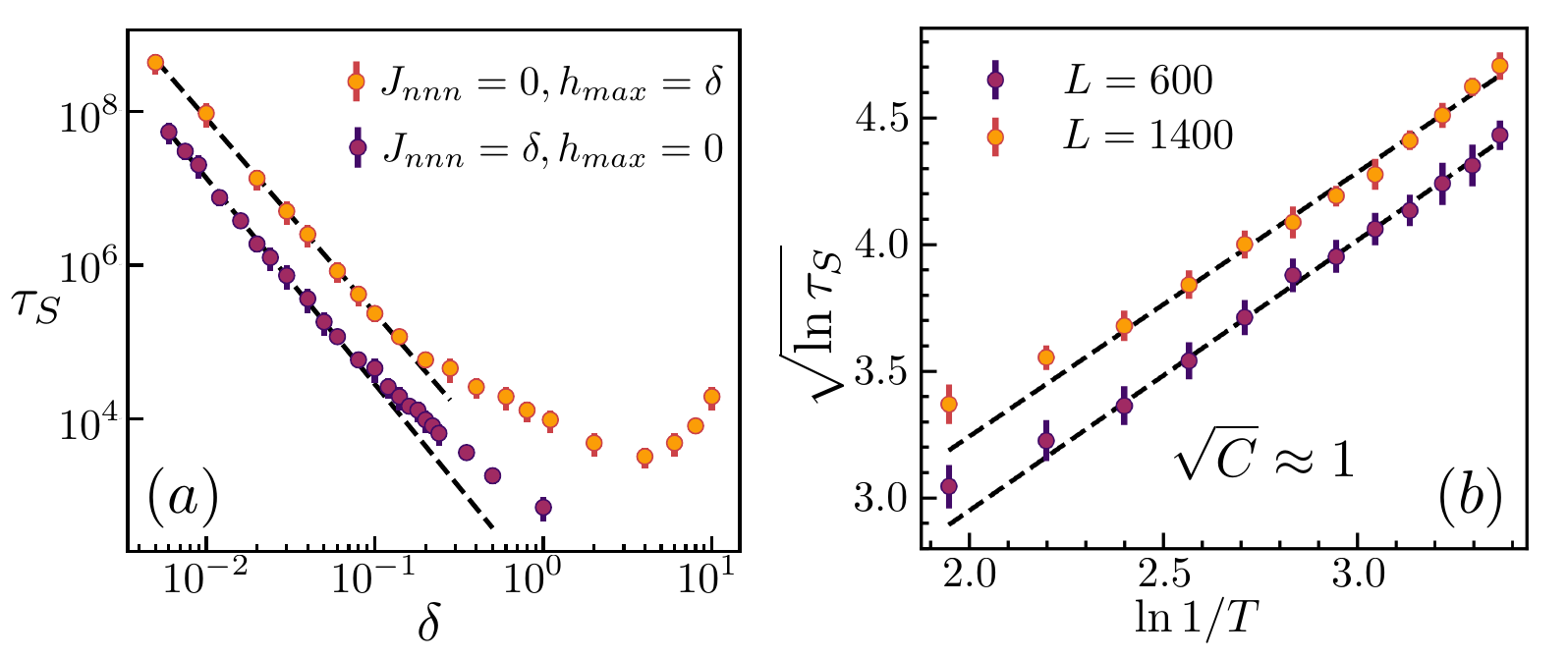}
	\caption{(a) Algebraic dependence of $\tau_S$ on chiral symmetry-breaking perturbation $\delta$ with parameter $1/T=1, L=1400$. Dashed black lines correspond to the algebraic fit $\delta^{-2.6}$.  (b) Delocalization time $\sqrt{\ln \tau_S}$ versus driving rate $\ln 1/T$ with parameter $h_{max}=1.1, J_{nnn}=0$. The numerical data fits well with a straight line of slope $\sqrt{C}\approx 1$, indicating $\tau_S$ behaves like $
\tau_{S} \sim e^{C[\ln (T^{-1} / g)]^{2}}$. Both figures use $J_0=1,J_{max}=1.6\pi,\delta h=1$ and 100 disorder average.}
		\label{fig:scaling_new}
\end{figure}
For weak breaking of the conservation law delocalization only occurs after a sufficiently long time scale. 
In Fig.~\ref{fig:scaling_new} (a), we show the dependence of the delocalization time  $\tau_S$ on different single-particle chiral symmetry-breaking perturbations of strength $\delta$. To quantify the dependence, one can extract the time $t_x$ at which the entanglement entropy crosses the threshold
$S_L(t_x)/(L/2)=x$. The prethermal lifetime $\tau_S$ is then defined as the average  $\tau_S = \langle t_x\rangle_x$ where the averaged is performed by using five different thresholds $x=s_0\pm\epsilon,s_0\pm\epsilon/2,s_0$ with the parameter values $s_0=0.11$ and $\epsilon=0.02$. 

It is worth noting that the driving frequency ($1/T=1$) is comparable to other energy scales of $H_0$. Hence, the prethermal localization cannot be predicted via a local effective Hamiltonian obtained from a high-frequency perturbative expansion~\cite{mori2021rigorous}. However, the prethermal localization can still be parametrically long-lived as it is protected by approximately conserved quantities $N_k$.
In the case of the small potential disorder $h_{max}=\delta$ (orange circles), the data fits well with a straight line in the log-log scale, suggesting that $\tau_S$ decreases algebraically as
$
\tau_S \propto \delta^{-\beta}
$
 with $\beta\approx2.6$. Interestingly, for $\delta>0.2$, a clear deviation from the scaling is observed and a turning point appears around $\delta\approx 5$, after which the prethermal lifetime instead increases monotonically. This suggests that beyond this point, instead of playing the role of a symmetry breaking perturbation, potential disorder stabilizes the localization and prolongs the prethermal lifetime.

Similarly, single-particle chiral symmetry can also be broken by introducing a static next-nearest-neighbor hopping  $J_{nnn}/2\sum_i(c_i^{\dagger}c_{i+2}+c_{i+2}^{\dagger}c_i)$ to $H_0$ and  approximately the same scaling exponent $\beta$ is extracted (purple circles). However, in contrast to potential disorder, long-range hopping cannot stabilize the localization, hence no turning point exists.

The prethermal lifetime $\tau_S$ also increases with the driving rate $1/T$, and we extract $\tau_S$ similar to above, but with  $s_0=0.13,\epsilon=0.01$.
We plot $\sqrt{\ln\tau_S}$ versus $\ln 1/T$ in Fig.~\ref{fig:scaling_new} (b) for two system sizes with nonzero potential disorder $h_{max}$. The numerical results fit well with a straight line of slope $\sqrt{C}\approx 1$ in the high frequency regime, indicating that the prethermal lifetime scales as
$
\tau_{S} \sim e^{C[\ln (T^{-1} / g)]^{2}}
$ with a local energy scale $g$
as proposed in Ref.~\cite{mori2021rigorous}. This scaling grows faster than any power law in $1/T$ but slower than exponential~\cite{mori2021rigorous,SM}.

\textit{Conclusion and Discussion.--}
We have identified a persistent localized steady state in a quasi-periodically driven multi-particle non-interacting system. The single-particle chiral symmetry and the resulting conservation law for multi-particle states account for this behavior.  One can also add higher powers of the drive operator $\hat{P}$ to the Hamiltonian, which leads to a long-range interaction and precludes the solvability of the model. However, as shown in the SM~\cite{SM}, such interaction cannot delocalize the system as it preserves the conservation law.

An analogous idea can be readily applied to systems hosting quantum many-body scars~\cite{turner2018weak,serbyn2021quantum} generated from, for instance, Hilbert space fragmentation~\cite{sala2020ergodicity,khemani2020localization,hudomal2020quantum,zhao2020quantum} or a spectrum generating algebra~\cite{moudgalya2018entanglement,schecter2019weak,shiraishi2017systematic}. As long as the fragmented structure is preserved by the drive, or the spectrum generating local ‘ladder’ operator~\cite{o2020tunnels}, which only couples low-entangled scared states, is modulated aperiodically, the system eventually will not thermalize.

The additional potential disorder, which breaks the single-particle chiral symmetry, spoils the conservation law and the system delocalizes after a long prethermal localized regime. Although the dependence of the prethermal lifetime on the driving frequency is well-understood for the TM drive, a microscopic description of the delocalization process and its dependence on the symmetry breaking is still missing and worth exploring in the future. In a related vein, a framework for the description of localization in the absence of a natural time-independent time evolution operator -- or perhaps a theory of the long-time convergence of the time evolution operators of aperiodic sequences -- would be highly desirable. 

\textit{Acknowledgements.--}
We acknowledge helpful discussion with Simon Lieu and Andrea Pizzi. We acknowledge support from the Imperial-TUM flagship partnership. HZ acknowledges support from a Doctoral-Program Fellowship of the German Academic Exchange Service (DAAD). This work was partly supported by the Deutsche Forschungsgemeinschaft under grants SFB 1143 (project-id 247310070) and the cluster of excellence ct.qmat (EXC 2147, project-id 390858490).

\bibliography{Loc_ref}
\bibliographystyle{unsrt}

\section{The chiral symmetry}
Here discuss details of the chiral symmetry. In the literature~\cite{PHSMBL,de2020anomalous}, instead of using the single-particle chiral symmetry generator $\hat{P}$, chiral symmetry for a fermionic system with Hamiltonian $H_{\mathrm{C}}=\sum_{i j} c_{i}^{\dagger} A_{i j} c_{j}$  is defined via the generator $\mathcal{S}=\mathcal{R}\times \mathcal{T}$, where $\mathcal{R}$ represents the 
particle-hole transformation $\mathcal{R}c_{i}^{\dagger}\mathcal{R}^{-1} =(-1)^ic_i$ and time reversal $\mathcal{T}c_{i}^{\dagger}\mathcal{T}=c_{i}^{\dagger},\mathcal{T}i\mathcal{T}^{-1}=-i$. The time reversal symmetry ensures that the matrix $\mathbf{A}$ is real and the particle-hole transformation takes the Hamiltonian to
\begin{eqnarray}
	\label{eq.PHtransformation}
	\begin{aligned}
		\mathcal{R}H_C\mathcal{R}^{-1} 
		&= \sum_{ij}(-1)^{i}c_iA_{ij}c_j^{\dagger}(-1)^j \\
		&=\sum_{ij}(-1)^{i+j+1}c_j^{\dagger}A_{ij}c_i+ \operatorname{tr}(\mathbf{A}) \\
		&=\sum_{i\neq j}(-1)^{i+j+1}c_i^{\dagger}A_{ij}c_j -\sum_{i}A_{ii}n_i+\operatorname{tr}(\mathbf{A}). \\
	\end{aligned}
\end{eqnarray}
As the initial Hamiltonian reads
\begin{eqnarray}
	H_{\mathrm{C}}=\sum_{i\neq j} c_{i}^{\dagger} A_{i j} c_{j}+\sum_{i}  A_{ii} n_i,
\end{eqnarray} 
by comparing it with Eq.~\ref{eq.PHtransformation}, we know that to preserve the chiral symmetry, the off-diagonal term needs to satisfy the condition
\begin{eqnarray}
	\label{eq.condition}
	(-1)^{i+j+1}A_{ij}=A_{ij},
\end{eqnarray}
and any diagonal terms in $\mathbf{A}$ has to be zero. Eq.~\ref{eq.condition} suggests that if $A_{ij}\neq 0$, $(-1)^{i+j+1}$ has to be 1, leading to the conclusion that $i+j$ is odd; otherwise, $A_{ij}$ is zero. These conditions guarantee that $\mathcal{S}H_{C}\mathcal{S}^{-1}=H_C$. Therefore, if $\ket{\epsilon_{\alpha}}$ is an eigenstate of $H_C$ with eigenenergy $\epsilon_{\alpha}$, $\mathcal{S}\ket{\epsilon_{\alpha}}$ is also an eigenstate with the same eigenenergy. However, these two eigenstates might not have the same particle number as the particle-hole transformation does not conserve it. 

For a single-particle state, Eq.~\ref{eq.condition} naturally leads to the anti-commutation relation $\{H_C,\hat{P}\}=0$ as $(-1)^i+(-1)^j$ in Eq.~\ref{eq.anticommutator} has to be zero. The Hamiltonian $H_0$ defined below Eq.~\ref{eq.fermi} in the main content is one special case with only nearest-neighbor hopping. $\hat{P}$ is unitary in the single-particle subspace with the property
\begin{eqnarray}
	\label{eq.identity}
	\hat{P}^2=\mathcal{I},
\end{eqnarray}
where $\mathcal{I}$ denotes the identity. The vanishing anti-commutator also  implies that for a given single-particle eigenstate $\ket{\epsilon_k}$ of eigenenergy $\epsilon_k$, $\hat{P}\ket{\epsilon_k}$ is another eigenstate of energy $-\epsilon_k$. Such a property sharply distinguishes the effect of $\mathcal{S}$ and $\hat{P}$.

In the multi-particle subspace, Eq.~\ref{eq.identity} is not true. Also, the anti-commutator
$\{H_C,\hat{P}\}= \hat{D},$ with $\hat{D}=2 \sum_{i jk} (-1)^{k}A_{i j} c_{i}^{\dagger}n_{k} c_{j}$, is generally non-vanishing for a multi-particle state. 
For example, consider a single body Fock state $\ket{\phi}_1 = c_{p}^{\dagger}|0\rangle$, we then have 
\begin{eqnarray}
	\hat{D} \ket{\phi}_1=\sum_{i j} A_{i j} c_{i}^{\dagger} \delta_{j p}\left(\sum_{k}(-1)^{k} n_{k}\right)|0\rangle,
\end{eqnarray}
where the number operator yeilds zero. However, for a two-body Fock state $\ket{\phi}_2 = c_{p}^{\dagger} c_{q}^{\dagger}|0\rangle$, one can verify that\begin{eqnarray}
	\label{eq.Deffect}
	\hat{D} \ket{\phi}_2=\sum_{i} A_{i p}(-1)^{q} c_{i}^{\dagger} c_{q}^{\dagger}|0\rangle\\-\sum_{i} A_{i q}(-1)^{p} c_{i}^{\dagger} c_{p}^{\dagger}|0\rangle,
\end{eqnarray}
which is a non-zero superposition of two-body Fock states. Therefore, $\{H_C,\hat{P}\}=0$ is valid only in the single particle subspace. For this reason we call $\hat{P}$ the `single-particle chiral symmetry generator' to distinguish it from the more standard chiral symmetry operator $\mathcal{S}$.

\section{Representation of $\hat{P}$ in the eigenbasis}
Here we will show how to expand $\hat{P} = \sum_i(-1)^in_i$ in the eigenbasis. 
By using the transformation $$c_i=\sum_{0<k\leq L/2}V_{ki}\gamma_k+\sum_{0<k\leq L/2}(-1)^iV_{ki}\tilde{\gamma}_k,$$ we obtain
\begin{eqnarray}
	\begin{aligned}
		\hat{P} = \sum_i(-1)^i \left( \sum_{0<q\leq L/2}V_{qi}^*\gamma^{\dagger}_q+\sum_{0<q\leq L/2}(-1)^i V_{qi}^*\tilde{\gamma}^{\dagger}_q \right)\\
		\times\left( \sum_{0<k\leq L/2}V_{ki}\gamma_k+\sum_{0<k\leq L/2}(-1)^iV_{ki}\tilde{\gamma}_k \right),
	\end{aligned}
\end{eqnarray}
where four combinations appear if we expand the product. However, two of them yield zero, for instance
\begin{eqnarray}
	\begin{aligned}
		\label{eq.zeroterm}
		&\sum_i(-1)^i \sum_{0<q\leq L/2}V_{qi}^*\gamma^{\dagger}_q \sum_{0<k\leq L/2}V_{ki}\gamma_k\\
		&= \sum_{0<q,k\leq L/2}\gamma^{\dagger}_q\gamma_k\sum_i(-1)^i    V_{ki}[V^{\dagger}]_{iq}\\
		&= \sum_{0<q,k\leq L/2}\gamma^{\dagger}_q\gamma_k\sum_i V_{\tilde{k}i}[V^{\dagger}]_{iq}\\
	\end{aligned}
\end{eqnarray}
where we define $V_{\tilde{k}i}=(-1)^iV_{ki}$. Suppose $V_{ki}$ represents the $k-$th eigenstate of energy $\epsilon_k$, then $V_{\tilde{k}i}$ corresponds to its paired state of energy $-\epsilon_k$. We also know $\sum_i V_{\tilde{k}i}[V^{\dagger}]_{iq}=\delta_{\tilde{k},q}$, which turns to be zero because $L/2<\tilde{k}\leq L$ whereas $0<{q}\leq L/2$. Therefore, Eq.~\ref{eq.zeroterm} gives zero. On the contrary, the following term is nonzero:
\begin{eqnarray}
	\begin{aligned}
		&\sum_i(-1)^i \sum_{0<q\leq L/2}V_{qi}^*\gamma^{\dagger}_q \sum_{0<k\leq L/2}V_{ki}\tilde{\gamma}_k (-1)^i \\
		&=\sum_{0<q,k\leq L/2}\gamma^{\dagger}_q\tilde{\gamma}_k\sum_i V_{ki}[V^{\dagger}]_{iq}\\
		&=\sum_{0<q,k\leq L/2}\gamma^{\dagger}_q\tilde{\gamma}_k\delta_{k,q}=\sum_{0<k\leq L/2}\gamma^{\dagger}_k\tilde{\gamma}_k.
	\end{aligned}
\end{eqnarray}
In the end, we obtain $\hat{P}_k=\gamma_{k}^{\dagger} \tilde{\gamma}_{k}+\tilde{\gamma}_{k}^{\dagger} \gamma_{k}$.

\section{Effect of $\hat{P}$ on different eigenstates}
As we discussed in the main content, $\hat{P}$ couples the paired single-particle excitations of  opposite energy. However, this will not be the case if we consider a multi-particle system. For example, we can consider the ground state $\left|\lambda_{G S}\right\rangle=\prod_{0<j<L/2} \tilde{\gamma}_{j}^{\dagger}|0\rangle$ where all negative energies are occupied. Operator $\hat{P}$ excites each of them forming a superposition as 
\begin{equation}
	\label{eq.coupling}
	\hat{P}\left|\lambda_{G S}\right\rangle=\sum_{0<i<L/2}(-1)^{L / 2-i} \prod_{0<j<L/2, j \neq i} \tilde{\gamma}_{j}^{\dagger} \gamma_{i}^{\dagger}|0\rangle,
\end{equation}
where the phase factor comes from anti-commutation relation for fermions.
This state is different from the state $\prod_{0<j<L/2} {\gamma}_{j}^{\dagger}|0\rangle$ of energy $\sum_{0<j<L/2} \epsilon_{j}$ as the symmetric counterpart of the ground state at the top edge of the multi-particle spectrum.
In addition, Eq.~\ref{eq.coupling} indicates that the ground state is only coupled with $L/2$ excited states of energy $-\sum_{0<j<L/2} \epsilon_{j}+2 \epsilon_{i}$, which is only a small fraction of the whole spectrum. Repeated application of $\hat{P}$ will couple more excitations. However, excitations at the same absolute energy cannot be simultaneously populated as the total occupation number $N_k$ for each $k$ must be conserved.

Another interesting two-body eigenstate is the one occupying both excitations at the same absolute energy:
\begin{eqnarray}
	\ket{\lambda_{k}} = \gamma_k^{\dagger}\tilde{\gamma}_k^{\dagger}\ket{0}.
\end{eqnarray}
This eigenstate has zero energy and does not couple to any other eigenstates via the operator $\hat{P}$ because $\hat{P}_k\ket{\lambda_k}=0$ holds. 

\section{Commutation between $N_k$ and $P_k$}
Here we show that the total number of excitations excitation $N_k=\gamma_{k}^{\dagger} \gamma_{k}+\tilde{\gamma}_{k}^{\dagger}\tilde{\gamma}_{k}$ at a fixed $k$ commutes with the drive $\hat{P}=\sum_{0<k\leq L/2} \gamma_{k}^{\dagger} \tilde{\gamma}_{k}+\tilde{\gamma}_{k}^{\dagger} \gamma_{k}$. As terms in different $k$-subspaces commute, we only need to check whether the following commutator
$
[P_k,\gamma_{k}^{\dagger} \gamma_{k}+\tilde{\gamma}_{k}^{\dagger}\tilde{\gamma}_{k}]
$ vanishes. It involves two contributions and the first one reads
\begin{eqnarray}
	\begin{aligned}
		\ &[P_k,\gamma_{k}^{\dagger} \gamma_{k}] = [\gamma_{k}^{\dagger} \tilde{\gamma}_{k}+\tilde{\gamma}_{k}^{\dagger} \gamma_{k},\gamma_{k}^{\dagger} \gamma_{k}]\\
		&=[\gamma_{k}^{\dagger} \tilde{\gamma}_{k},\gamma_{k}^{\dagger} \gamma_{k}]+[\tilde{\gamma}_{k}^{\dagger} \gamma_{k},\gamma_{k}^{\dagger} \gamma_{k}]\\
		&=\gamma_{k}^{\dagger} \tilde{\gamma}_{k}\gamma_{k}^{\dagger} \gamma_{k}-\gamma_{k}^{\dagger} \gamma_{k} \gamma_{k}^{\dagger} \tilde{\gamma}_{k}    +[\tilde{\gamma}_{k}^{\dagger} \gamma_{k},\gamma_{k}^{\dagger} \gamma_{k}]\\
		&=0-\gamma_{k}^{\dagger}(1-  \gamma_{k}^{\dagger} \gamma_{k})\tilde{\gamma}_{k}    +[\tilde{\gamma}_{k}^{\dagger} \gamma_{k},\gamma_{k}^{\dagger} \gamma_{k}]\\
		&=-\gamma_k^{\dagger}\tilde{\gamma}_k    +[\tilde{\gamma}_{k}^{\dagger} \gamma_{k},\gamma_{k}^{\dagger} \gamma_{k}],\\
	\end{aligned}
\end{eqnarray}
where we use $(\gamma_k^{\dagger})^2=0$. In the end we get
\begin{eqnarray}
	\label{eq.onecontribution}
	[P_k,\gamma_{k}^{\dagger} \gamma_{k}] = -\gamma_k^{\dagger}\tilde{\gamma}_k   +\tilde{\gamma}_k^{\dagger} \gamma_k.
\end{eqnarray}

The second contribution can be obtained similarly as $
[P_k,\tilde{\gamma}_{k}^{\dagger}\tilde{\gamma}_{k}] = -\tilde{\gamma}_k^{\dagger}{\gamma}_k   +{\gamma}_k^{\dagger} \tilde{\gamma}_k.
$ It cancels Eq.\ref{eq.onecontribution}, hence we obtain $
[P_k,\gamma_{k}^{\dagger} \gamma_{k}+\tilde{\gamma}_{k}^{\dagger}\tilde{\gamma}_{k}]=0.
$

\section{Localization in the long-time limit}
Using a pseudo-spin representation, one can rewrite the elementary time evolution operator for each subspace labeled by $k$ in the form 
\begin{equation}
	\begin{array}{l}
		U^k_{\pm}=\exp \left[-i T E_{k}\left(\sin \gamma_{k} \sigma^k_{z}\pm\cos \gamma_{k} \sigma^k_{x}\right)\right],
	\end{array}
\end{equation}
with $E_{k}=\sqrt{\epsilon_{k}^{2}+\delta h^{2}},$ and $\sin \gamma_{k}=\epsilon_{k} / E_{k}, \cos \gamma_{k}=\delta h / E_{k}$. The unitary operator recursively determined by Eq.~\ref{eq.recursive} converges to $U^k_{\infty}=\exp\big(-i\alpha^k_{\infty}\sigma_z\big)$ in the limit $n\to\infty$ with a nonsaturating phase factor $\alpha^k_n$~\cite{nandy2017aperiodically}. Hence, in the asymptotic limit it is diagonal in the $z$ direction, and $U^k_{\infty}$ can directly be diagonalized by $\ket{\uparrow}_k,\ket{\downarrow}_k$, the same eigenstates of the static Hamiltonian $H_{\mathrm{0}}$. 

\section{Comparison between different fits}
\begin{figure}
	\centering
	\includegraphics[width=0.48\linewidth]{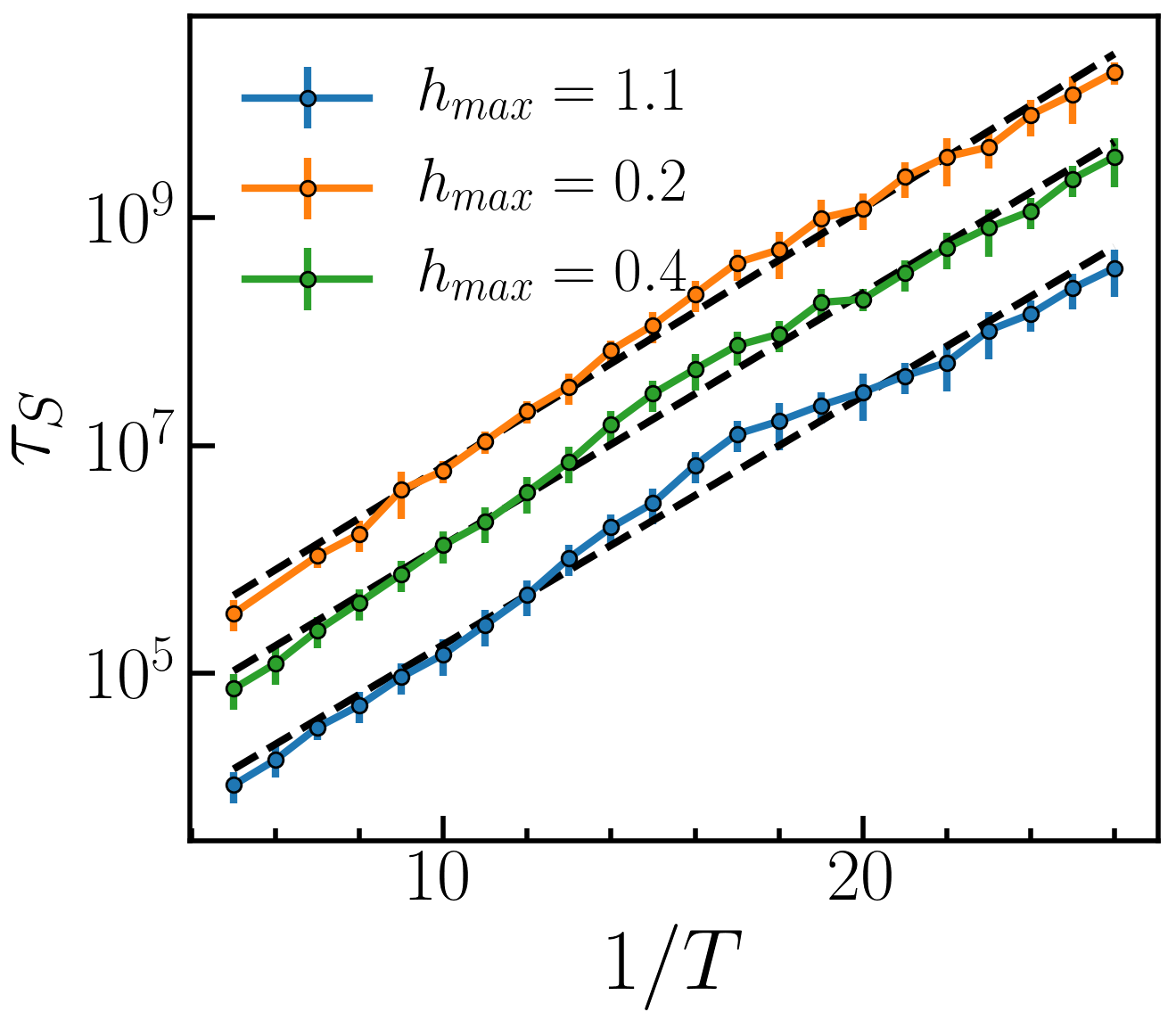}
	\includegraphics[width=0.49\linewidth]{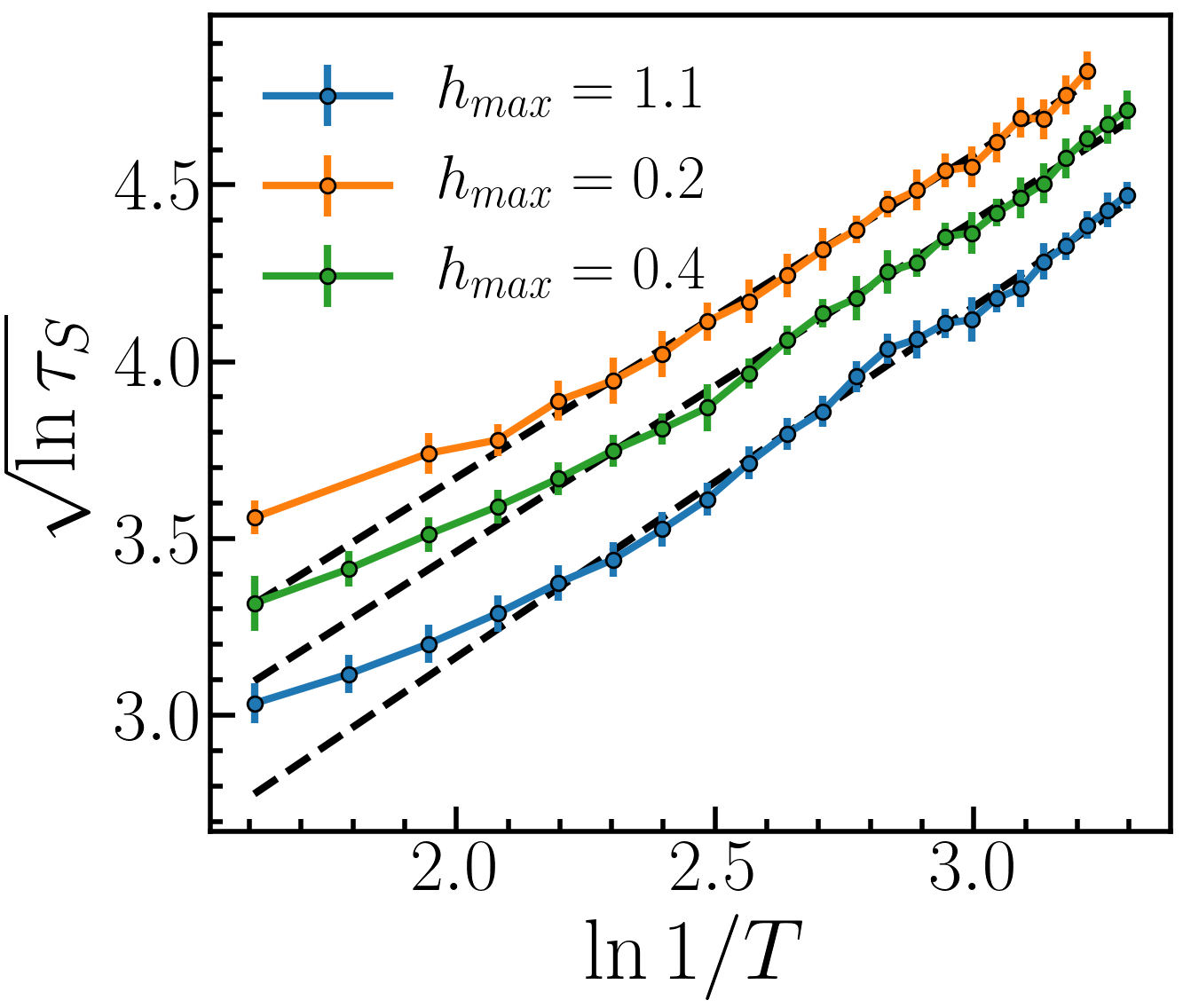}
	\caption{ $J_0=1,J_{max}=1.6\pi,\delta h=1,L=600$.  }
	\label{fig:scaling_com}
\end{figure}
Here we extract the delocalization time by averaging the time $t$ such that $S_L(t)/(L/2)=0.2\pm0.04,0.2,0.2\pm0.02$, and compare different fitting methods. 
In the left panel of Fig.~\ref{fig:scaling_com}, we plot the numerical data in the log scale and fit it with the exponential scaling
\begin{eqnarray}
	\tau_S\sim \exp(\alpha T^{-1}).
\end{eqnarray}
The slope of the linear fit corresponds to the scaling exponent $\alpha\approx0.5$. The numerical data curve down from a straight dashed line, suggesting that the heating time is shorter than exponential. As a comparison, in the right panel we perform the fitting only for $1/T>7$, and the data fit well with 
\begin{equation}
	\tau_{S} \sim e^{C[\ln (T^{-1} / g)]^{2}},
\end{equation}
with $\sqrt{C}\approx 1$. Both the scaling parameters $\sqrt{C}$ and $\alpha$ do not depend on $h_{max}$.

\section{Interaction effects}
\begin{figure}
	\centering
	\includegraphics[width=0.49\linewidth]{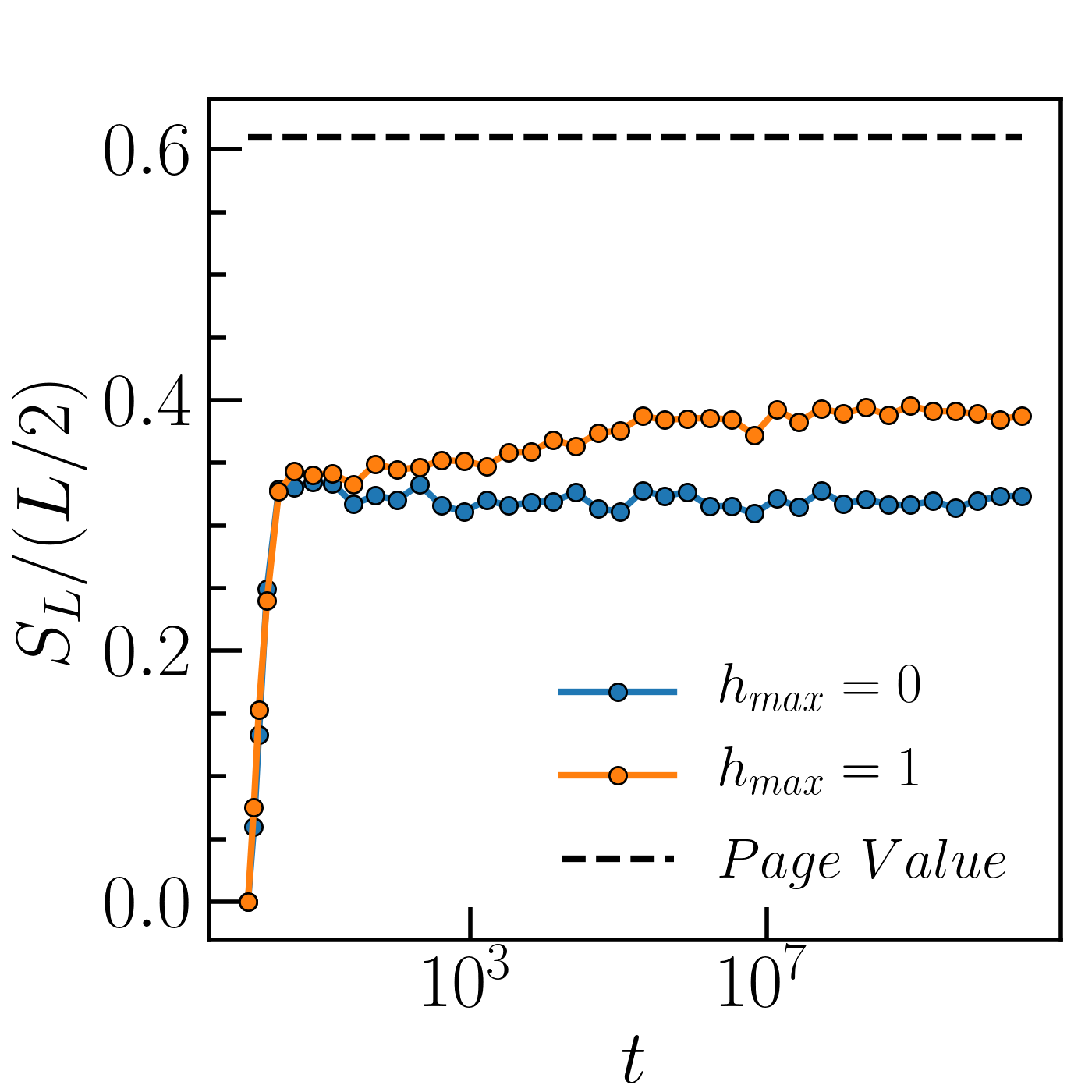}
	\includegraphics[width=0.49\linewidth]{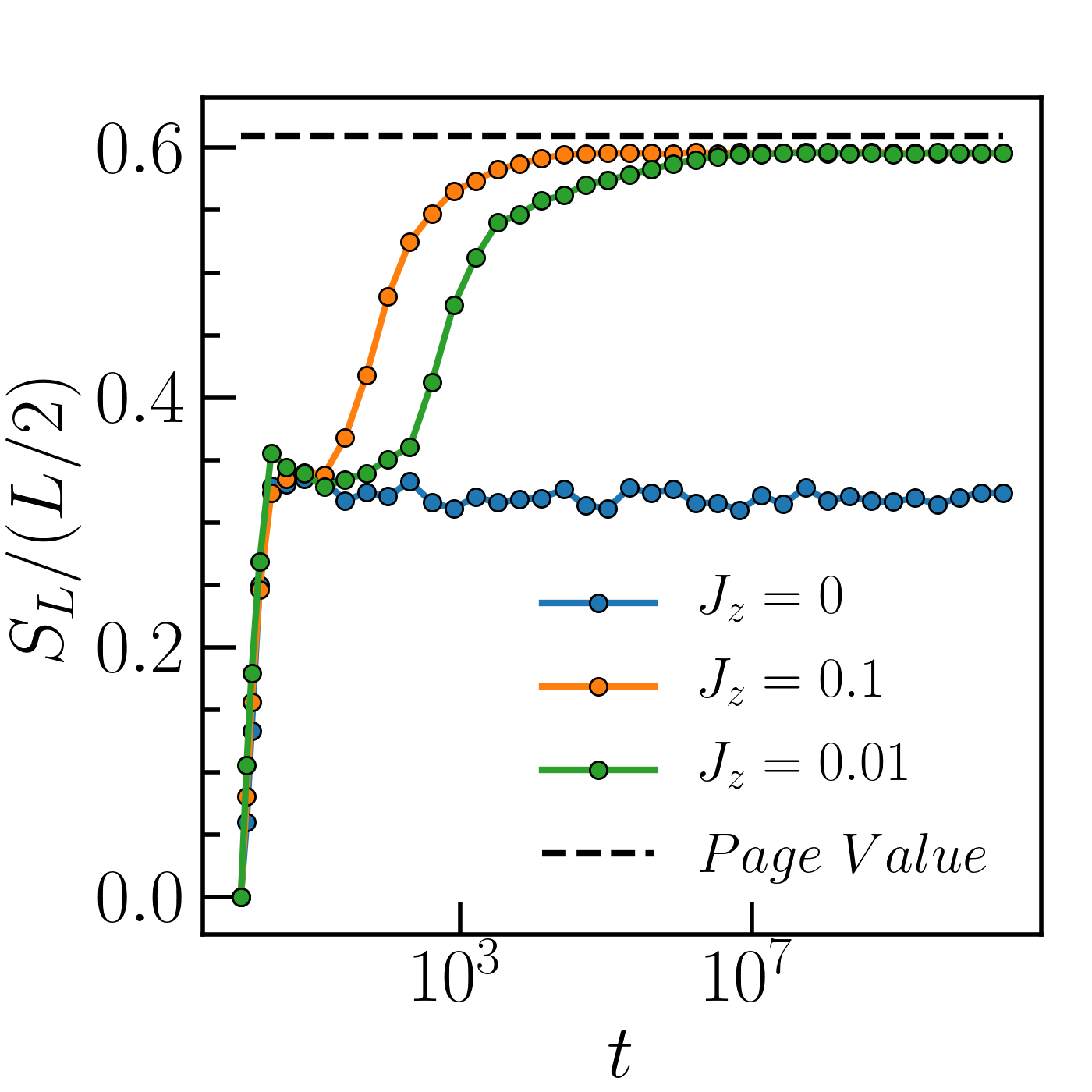}
	\caption{ $J_0=1,J_{max}=1.6\pi,\delta h=1,T^{-1}=10$.  }
	\label{fig:dynamics_nonintegrable}
\end{figure}

As shown in Fig.~\ref{fig:entanglement} of the main text, the final entanglement plateau does not reach the Page value. It indicates that although the system thermalizes, the final state is not the infinite temperature state because the system is non-interacting. 
Here we introduce the interaction in z direction as a perturbation to explicitly break the integrability. The Hamiltonian then reads
\begin{eqnarray}
	\begin{aligned}
		H_{\pm} &=H_0+ \frac{1}{2}\sum_{i=1}^{L} h_i^{\pm} \sigma_i^z, \\
		H_0 &= \frac{1}{4}\sum_{i=1}^{L-1}J_i\big(\sigma_i^x\sigma_{i+1}^x+\sigma_i^{y}\sigma_{i+1}^y\big)+J_z\sigma_i^z\sigma_{i+1}^z, 
	\end{aligned}
	\label{eq.Hamiltonian_Jz}
\end{eqnarray}
and the simulation is performed via exact diagonalization for system size $L=12$. The result is plotted in Fig.~\ref{fig:dynamics_nonintegrable}.

First, let us focus on the non-interacting case (blue data in the left panel) with $J_z=0$ and $h_{max}=0$ where the entanglement quickly saturates after a transient  relaxation. However, we notice this value still depends on system size indicating that the simulation is limited by the finite size effect. Nonzero $h_{max}$ (orange dots) delocalizes the system and entanglement increases and saturates at a value smaller than the Page value $S_L=(L\log2-1)/2$ (black dashed line) corresponding to the infinite temperature state.

In contrast, as shown in the right panel, even for $h_{max}=0$, the interaction $J_z\sum_{i}\sigma_i^z\sigma_{i+1}^z$ delocalizes the system. This happens because $J_z\sum_{i}\sigma_i^z\sigma_{i+1}^z$ breaks the conservation law $N_k$, even though it is invariant under the chiral transformation $\mathcal{S}$ up to a constant. As this perturbation also breaks 
the integrability, the final saturation is close to the Page value, suggesting  eventual thermalization to  infinite temperature.

\begin{figure}[h]
	\centering	\includegraphics[width=0.49\linewidth]{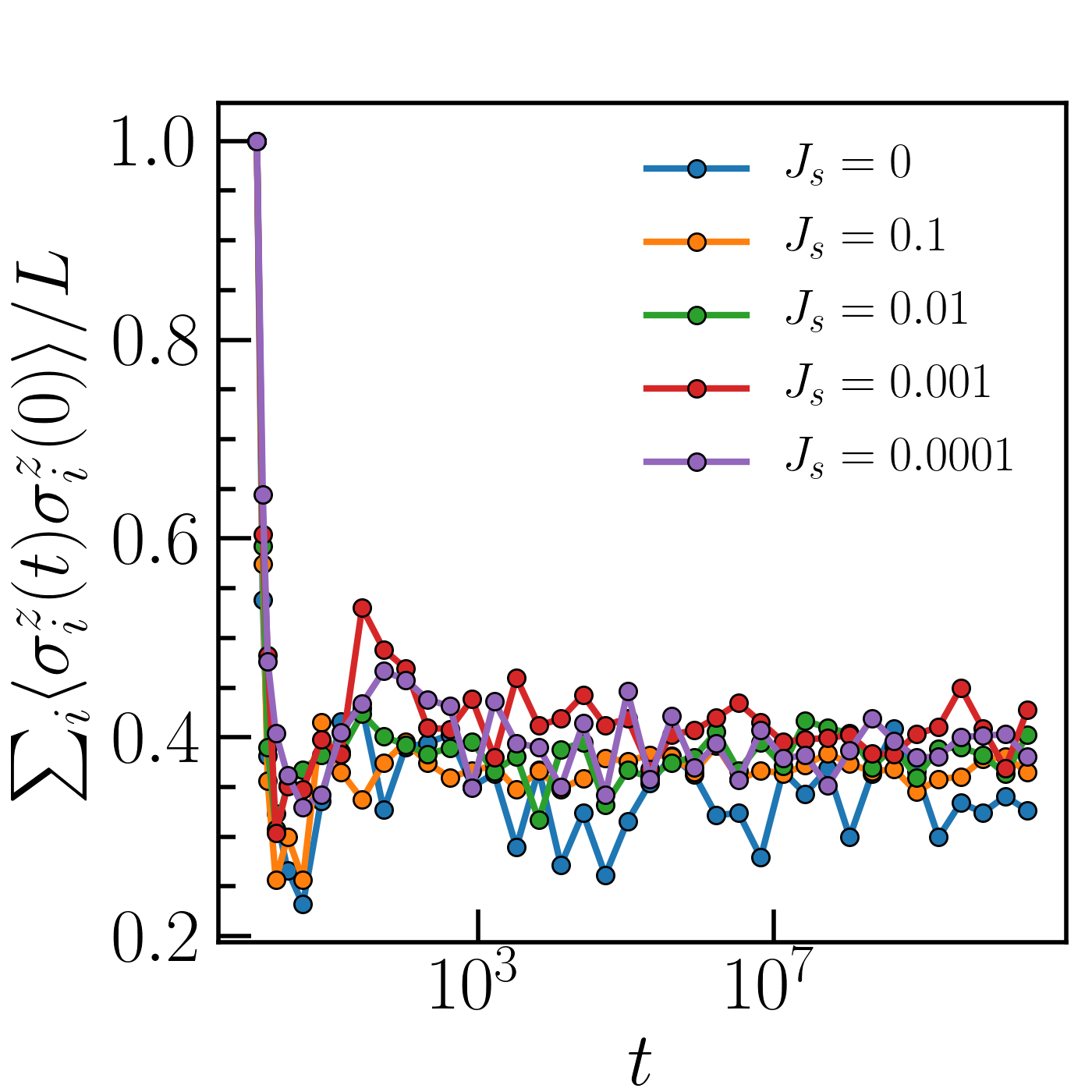}
	\includegraphics[width=0.49\linewidth]{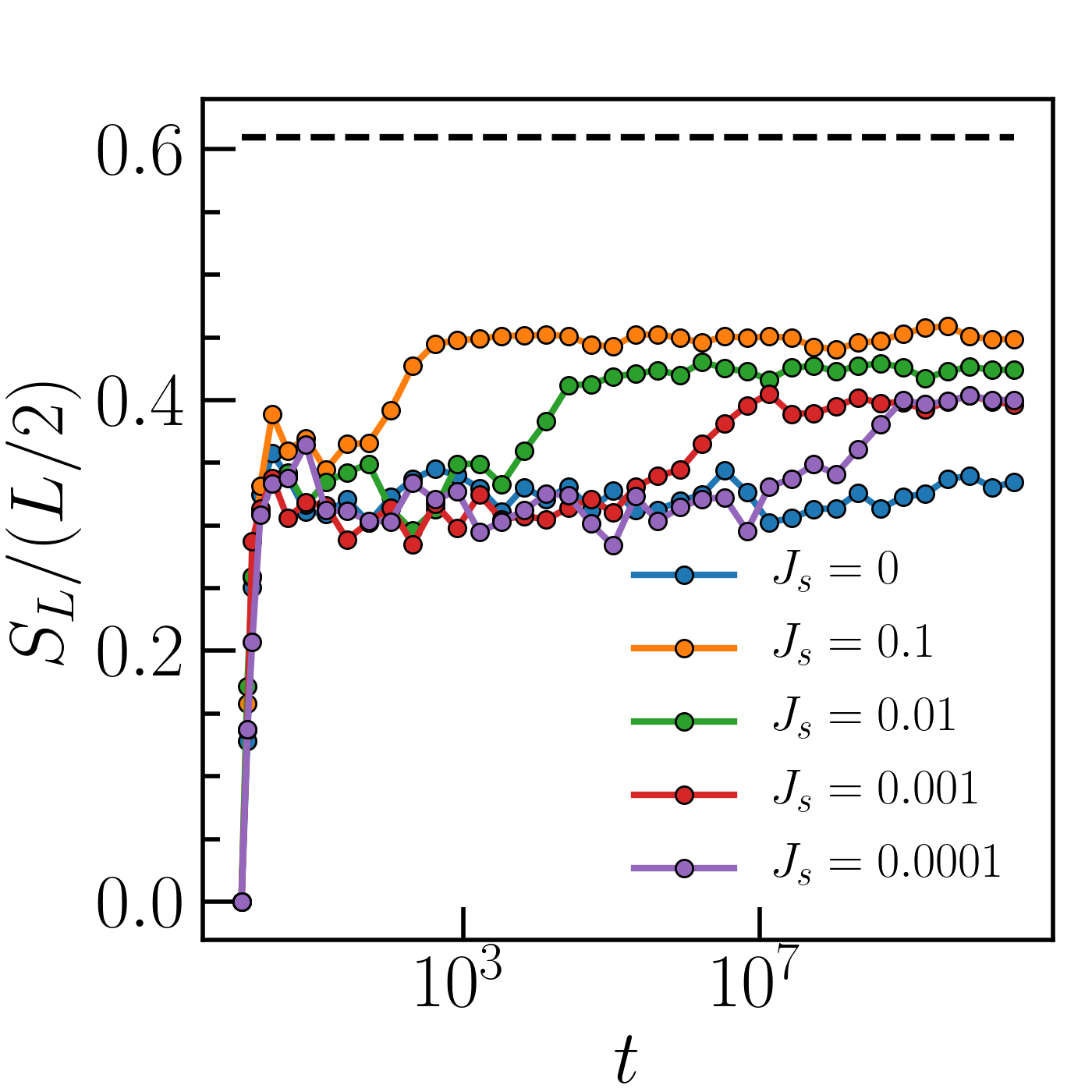}
	\caption{ $J_0=1,J_{max}=20,\delta h=1,h_{max}=0,1/T=10$.  }
	\label{fig:dynamics_Js}
\end{figure}

Alternatively, we also consider interactions which preserve the conservation of $N_k$, for instance $\hat{P}^2$. In terms of spin operators, up to a constant, this term can be expressed as 
\begin{eqnarray}
	H_s = \frac{J_s}{4}\sum_{i,j}(-1)^{i+j}\sigma_i^z\sigma_j^z,
\end{eqnarray}
which is an infinitely long-range interaction in $z$ direction.
The full driving now reads
\begin{eqnarray}
	\begin{aligned}
		H_{\pm} &=H_0+ \frac{1}{2}\sum_{i=1}^{L} h_i^{\pm} \sigma_i^z \pm H_s.
	\end{aligned}
	\label{eq.Hamiltonian_Js}
\end{eqnarray}
As shown in Fig.~\ref{fig:dynamics_Js} left panel, starting from the domain-wall initial state $\ket{\uparrow\dots\uparrow\downarrow\dots\downarrow}$ with zero total magnetization, the system maintains the memory of the initial state for small values of $J_s$, suggesting that the system is localized in the presence of the conservation-preserving interaction. However, as shown in the right panel, non-vanishing $J_s$ tends to generate more entanglement than the non-interacting case. The steady state does not heat up to infinite temperature because the system is localized and ${N}_k$ is conserved. 

The comparison between different types of interactions highlights that the persistent localization is protected by the single-particle chiral symmetry generated by $\hat{P}$, instead of the chiral symmetry generated by $\mathcal{S}$.

\end{document}